\documentclass[twocolumn,prb]{revtex4}
\usepackage{epsfig}
\usepackage{epstopdf}
\usepackage{amsmath,amsfonts,paralist}
\usepackage{amssymb}
\usepackage{graphicx}
\usepackage{mathrsfs} 
\usepackage{color}
\usepackage{natbib,hyperref,url,textcase}

\usepackage{dcolumn}
\usepackage{bm}     
\usepackage{dsfont,float}
\usepackage{upgreek}
\usepackage{subfigure}

\newcommand{\bG     }{\mbox{\boldmath$G$}}
\newcommand{\bphi     }{\mbox{\boldmath$\phi$}}

\begin{document}

\title{The renormalization flow of the hierarchical Anderson model at weak disorder}

\author{F. L. Metz$^{1}$, L. Leuzzi$^{1,2}$, G. Parisi$^{1,2,3}$}
\affiliation{$^1$ Dip. Fisica, Universit{\`a} {\em La Sapienza}, Piazzale
  A. Moro 2, I-00185, Rome, Italy \\ $^2$ IPCF-CNR, UOS Roma {\em
    Kerberos}, Universit\`a {\em La Sapienza}, P. le A. Moro 2, I-00185,
  Rome, Italy \\ $^3$ INFN, Piazzale A. Moro 2, 00185, Rome, Italy}
\date{\today}  

\begin{abstract}
We study the flow of the renormalized model parameters obtained
from a sequence of simple transformations of the 1D Anderson model 
with long-range hierarchical hopping.
Combining numerical results with a perturbative approach for
the flow equations, we identify three qualitatively different regimes at
weak disorder.
For a sufficiently fast decay of the hopping energy, the Cauchy distribution
is the only stable fixed-point of the flow equations, 
whereas for sufficiently slowly decaying hopping energy the renormalized parameters
flow to a delta peak fixed-point distribution. In an intermediate range of the hopping decay, both fixed-point distributions are
stable and the stationary solution is determined by the initial configuration of 
the random parameters. We present results for the critical decay of the
hopping energy separating the different regimes.
\end{abstract}
\maketitle

\section{Introduction} 

The localization of a quantum particle in the
presence of a random potential remains a very
active topic in condensed matter physics. \cite{Evers2008} The prototypical model
of localization is the Anderson tight-binding
model with short-ranged, nearest-neighbour hoppings on an hypercubic lattice. \cite{Anderson58} 
Such model undergoes a transition between extended and localized
wavefunctions for sufficiently high spatial dimensions. \cite{Gang4} In the
context of random matrices, a 
popular class of models is represented by the Wigner ensemble \cite{Wigner}, where the 
matrix elements
are Gaussian distributed random variables and 
the fully-connected infinite-range character of the hopping energies prevents the wavefunctions to become localized. 
To study an intermediate situation between
these two fundamental models, one requires a hopping energy
decaying slowly as a function of the intersite distance.

The hierarchical Anderson model (HAM), introduced originally
by Bovier, \cite{Bovier}
is a tight-binding model
with on-site disorder and  hopping energies organized in an
hierarchical block structure. The hopping energy falls off as
a power-law for large intersite distances, allowing to interpolate
smoothly between models with short-range and infinite-range 
hopping energy. 
Hierarchical models have a long tradition in statistical
physics, which goes back to 
Dyson, \cite{Dyson1969} and they constitute
an approximate route to study the behavior
of models defined in terms of the standard
short-range Laplacian on the hypercubic lattice, such as
the classical random walk \cite{Mezard,Molchanov1996} and interacting 
spin systems. \cite{Meurice2007,Jacopo}
Besides that, hierarchical
models are conveniently designed such that
they preserve their structure under renormalization
transformations, \cite{Bovier,Meurice2007,Baker72} being amenable to an
exact and thorough analysis.


Contrary to the rigorous results established
for the density of states (DOS), \cite{Molchanov1996,Kr07,Kritch07,Kr08,Muller2012}
less work has been devoted to the study of
the nature of the eigenstates of the HAM.
In a recent paper, \cite{Metz2013} the authors have shown the
existence of an extended phase for a sufficiently slow decay
of the hopping energy, 
in contrast
to a previous conjecture stating that all states should be
localized. \cite{MonthusGarel} The results are based
on a renormalization procedure for the resolvent 
matrix, which allows to compute numerically the inverse participation
ratio (IPR) for extremely large system sizes. 
In the present work we address the problem of identifying and analyzing 
the qualitative change in the fixed-point distribution of
the flow equations corresponding to the onset of a  localization transition in the HAM.
The stationary solution of the flow equations 
has been studied so far only in the strong disorder regime, \cite{MonthusGarel} where
the Cauchy distribution is the only fixed-point and all
eigenstates are localized. 

Here we complement the work presented in Ref. [\onlinecite{Metz2013}] by
studying the flow equations for the renormalized random
potentials (RRP) resulting from the consecutive elimination
of the degrees of freedom of the resolvent matrix via a simple
change of integration variables.
We focus on  the stability of the fixed-point distribution
of the RRP at the band edge of the pure spectrum, when a small amount of on-site disorder is introduced.  
The motivation for studying this specific situation is twofold. First, the 
problem concerns the survival
of the band edge extended  wavefunction of the pure model 
in the presence of weak disorder.
Second, this
is the energy range where the integrated DOS of the HAM 
has a similar behaviour as that exhibited by
short-range systems in finite dimensions, and we expect that our work provides
further insights on the behavior of the latter.

By means of the numerical solution of the flow equations, combined
with a perturbative approach, we show that three distinct regimes
emerge at weak disorder. The distribution
of the RRP flows to a Cauchy
distribution  fixed-point provided the hopping energy decays sufficiently fast.
For a sufficiently slow decay of the hopping energy, the fluctuations
of the RRP vanish exponentially and
the flow converges to a delta peak distribution. In an intermediate
region of the hopping energy decay, our numerical results
suggest that both fixed-point distributions are stable and
the asymptotic behaviour depends on the specific microscopic configuration
of the on-site disorder. We quantify the
basin of attraction of both solutions by computing numerically
the fraction of the flow that has evolved to a delta peak
distribution. As it will be explained later, our conclusions
are valid for finite, but very large system sizes.

The paper is organized as follows. We define
the model in section \ref{secmodel}. The renormalization
procedure used in deriving the flow equations
is explained in section \ref{secrenorm}, while a perturbative
expansion of these equations is presented in section  \ref{secperturb}.
The discussion of the  numerical results, guided
by the outcome of the perturbative approach, is left for section \ref{numsec}.
Finally, we present some final remarks in the last section.

\section{The hierarchical Anderson model} \label{secmodel}

Tight-binding models constitute 
the simplest lattice models to study the diffusion of a
quantum particle in the presence of a spatially random
potential. \cite{Anderson58} We consider a one-dimensional chain
of unity lattice spacing composed of $L=2^{N}$ sites $i=1,\dots,L$, with 
random potentials $\{  \varepsilon_{i} \}_{i=1,\dots,L}$ drawn from
a distribution $p(\varepsilon)$. At this
stage there is no need to specify $p(\varepsilon)$ and we keep the model definitions as general as possible. 
In the hierarchical Anderson model the kinetic energy is given in terms of
a hierarchical Laplacian. \cite{Bovier}
Inspired by the original work of Dyson, \cite{Dyson1969} we define the Hamiltonian as follows
\begin{eqnarray}
\mathcal{H}_{N} &=& \sum_{i=1}^{2^{N}} \varepsilon_{i} \mid i \rangle \langle i \mid
\label{hamilt}
\\
&+& J \sum_{p=1}^{N} V_{p} \sum_{r=1}^{2^{N-p}} 
\sum_{i\neq j}^{1,2^{p}}  \mid (r-1)2^p + i \rangle  \langle (r-1)2^p + j \mid \,,
\nonumber
\end{eqnarray}
where $| i \rangle $ is the canonical site basis.  

The hierarchy of hopping energies has a total number 
of $N$ levels, where $p=1$ and $p=N$ denote, respectively, the
lowest and the highest level of the hierarchy.
At each level the system is divided into $2^{N-p}$ distinct 
blocks, each of which contains $2^p$ sites. 
The hopping between any two sites within a single block of level $p$ has energy $J V_p$, while 
the hopping between sites in two different blocks is determined by levels higher 
in the hierarchy and has energy $t_p = J \sum_{n=p}^{N} V_n$, where
$J$ sets the scale of energy.
A schematic representation of this hierarchical block
structure of the kinetic energy is presented in Ref. [\onlinecite{Metz2013}].

Distinctly from the case of ultrametric random matrices, \cite{Ossipov2009} where
the hierarchical structure is encoded in
the choice of variances for the Gaussian distributed hoppings
between the sites, here $\{ V_{p} \}_{p=1,\dots,N}$ are 
non-random parameters.
We choose them to decay as a function
of the level index according to $V_p = 2^{-\alpha(p-1)}$, where
$\alpha > 1$ controls the
speed of the decay. This restriction on 
$\alpha$ ensures that, in the absence of disorder, the support of the DOS 
is bounded for $L \rightarrow \infty$ (see Eq. (\ref{accpoint})).
For $N \gg 1$, the magnitude of the hopping energy between two sites
separated by a distance of $O(L)$ scales as 
$O(1/L^{\alpha})$, exhibiting the same long-distance
behaviour as a tight-binding model with size $L$ and hopping
energy decaying as a power $\alpha$ of the intersite distance. \cite{Yeung87,Rodriguez2000,Rodriguez2003,Malyshev2004,Moura2005}

For $p(\varepsilon) = \delta(\varepsilon)$, the eigenvalues and eigenvectors
of the Hamiltonian (\ref{hamilt}) can be
computed analytically \cite{Bovier,MonthusGarel} and
the average DOS reads 
\begin{equation}
\rho_{{\rm pure}}(E) = \sum_{p=1}^{\infty} \frac{1}{2^{p}} \delta(E - E^{{\rm pure}}_{p-1})\,,
\label{rhopure}
\end{equation}
where 
\begin{equation}
E^{{\rm pure}}_{p} =- \frac{J}{\left( 1 - 2^{-\alpha} \right)} 
+ 2 J \left[ \frac{1 - 2^{-(\alpha-1)p}}{  1 - 2^{-(\alpha-1)}}   \right] \,.
\label{energpure}
\end{equation}
The average DOS is a series of Dirac delta
peaks, which may be interpreted as arising
from flat bands. Each peak in $\rho_{{\rm pure}}(E)$ corresponds to a 
level of the hierarchy and the factor $2^{-p}$
comes from the degeneracy induced by the symmetry
between the blocks at each level.
The delta peaks accumulate 
at the upper spectral edge 
\begin{equation}
E^{{\rm pure}}_{\infty} = - \frac{J}{\left( 1 - 2^{-\alpha} \right)}  
+ \frac{2 J}{\left( 1 - 2^{1-\alpha} \right)}\, ,
\label{accpoint}
\end{equation}
where $\alpha > 1$ ensures that 
$E^{{\rm pure}}_{\infty} = \lim_{p \rightarrow \infty} E^{{\rm pure}}_{p}  < \infty$.
The IPR of a normalized eigenstate $| \psi \rangle$ is
defined as 
\begin{equation}
I = \sum_{i=1}^{L}  \langle i | \psi \rangle^4 \,.
\end{equation}
In the pure model the IPR of the eigenstate at $E = E^{{\rm pure}}_{\infty}$  
scales as  $I = 1/L$, corresponding to an extended wavefunction.

%
%

The integrated density of
states  of the
pure HAM is \cite{Molchanov1996}
\begin{equation}
\mathcal{N}(E^{{\rm pure}}_{p}) = \sum_{\ell=1}^{p} 2^{-\ell}
= 1 - C \left( E^{{\rm pure}}_{\infty} - E^{{\rm pure}}_{p} \right)^{d_s/2}
\end{equation} 
where
\begin{eqnarray}
\nonumber
C &=& \left( E^{{\rm pure}}_{\infty} - E^{{\rm pure}}_{0} \right)^{-d_s/2}
\\
\nonumber
d_s &=& \frac{2}{\alpha-1}.
\end{eqnarray}
Therefore, close to the upper spectral edge $E^{{\rm pure}}_{\infty}$, the integrated
DOS exhibits the asymptotic 
behaviour \cite{Molchanov1996,Kr07,Kr08,Muller2012}
\begin{equation}
\nonumber
1 - \mathcal{N}(E) \sim  \left( E^{{\rm pure}}_{\infty} - E \right)^{d_s/2} \ .
\end{equation}  
The number $d_s$ is the {\it spectral dimension} \cite{Kr07,Kr08,Muller2012} and its definition
is motivated by noting that the same band edge asymptotics of the integrated DOS
is observed in the case of the short-range Laplacian on an hypercubic lattice, for which the 
spectral and the spatial dimension coincide.

The integrated DOS of the HAM also presents the same band edge asymptotics
as that found in the pure one-dimensional tight-binding model with power-law decaying 
hopping energy, with an exponent in the range $1 < \alpha < 2$. \cite{Balagurov2004}
Therefore, the integrated DOS and the IPR of models with long-range
hopping energies exhibit, in the neighbourhood of the upper spectral edge, the same behaviour
as that found in the short-range Laplacian on spatial dimension $D$, as
long as $\alpha$ is chosen such that $d_s=D.$ \cite{Molchanov1996}
Consistent with that, the HAM undergoes a localization transition
close to $E^{{\rm pure}}_{\infty},$ \cite{Metz2013} and
this is the interesting region to study the flow of the
renormalized parameters.

\section{The renormalization flow equations} \label{secrenorm}

In this section we discuss the main ideas involved
in the derivation of the equations
describing the flow of the RRP. 
The central object of our approach is the resolvent
matrix 
\begin{equation}
\nonumber
\bG^{(N)} = \frac{1}{z - \mathcal{H}_{N}}
\end{equation}
 of
the HAM with $N$ levels, where $z = E - i \eta$ and $\eta > 0$
is a regularizer. The resolvent elements in the site basis can be represented in terms of 
Gaussian integrals according to 
\begin{equation}
G_{ij}^{(N)} = i \frac{\int d \bphi^{(N)}  \, \phi_{i} \phi_{j}
\exp{   \left[    S^{(N)}( \bphi| \mu_{1,\dots,2^{N}}, V_{1,\dots,N}   ) \right] }}
{\int  d \bphi^{(N)}
\exp{\left[ S^{(N)}( \bphi| \mu_{1,\dots,2^{N}}, V_{1,\dots,N} ) \right] }} \,,
\label{Greendef}
\end{equation}  
where $d \bphi^{(N)} \equiv \prod_{i=1}^{2^{N}} \phi_i$ and
\begin{eqnarray}
S^{(N)}(  \bphi| \mu_{1,\dots,2^{N}}, V_{1,\dots,N} )  &=& \frac{i}{2} \sum_{i=1}^{2^N}  \mu_i \phi_{i}^{2} 
\label{actioneq}
\\ 
&+& J \, W^{(N)}( \phi_{1,\dots,2^{N}}, V_{1,\dots,N} ) . 
 \nonumber 
\end{eqnarray}
We have introduced the shorthand notation $x_{1,\dots,\mathcal{A}} \equiv x_1,\dots,x_{\mathcal{A}}$ to
represent sets of variables.
The local parameters 
\begin{equation}
\mu_i = \varepsilon_{i} - J \sum_{p=1}^{N} V_p  - z
\end{equation}
include the random potentials, while $W^{(N)}$ encodes
the hierarchical hopping contribution
\begin{eqnarray}
W^{(N)}( \phi_{1,\dots,2^{N}}, V_{1,\dots,N} ) =     \frac{i}{2} \sum_{p=1}^{N} 
V_{p} \sum_{r=1}^{2^{N-p}} \left( \sum_{j=1}^{2^{p}} \phi_{(r-1)2^{p} +j}   \right)^2   
\nonumber 
\end{eqnarray} 

The essential idea consists in obtaining a recursion relation between the resolvent of
a system with $2^{N}$ sites and the resolvent of a system
with $2^{N-1}$ sites, with renormalized model parameters. 
The change of integration variables 
$\psi^{\pm}_{i} = \frac{1}{\sqrt{2}} (\phi_{2i-1} \pm \phi_{2i} )$ ($ i=1,\dots,2^{N-1}$)
in Eq. (\ref{Greendef})
allows to calculate explicitly the integrals
over $\{ \psi^{-}_{i} \}_{i=1,\dots,2^{N-1}}$, halving 
the number of degrees of freedom.
The function $S^{(N-1)}$, following from this integration, has
the same formal structure as Eq. (\ref{actioneq}), 
reflecting the invariance of the Hamiltonian under a
renormalization transformation. 
After applying this change of variables $\ell$ times in a consecutive
way, we obtain an expression for the 
resolvent elements $G_{ij}^{(N-\ell)}$ which is formally the
same as Eq. (\ref{Greendef}), 
but depends on the function 
\begin{eqnarray}
S^{(N-\ell)}(\bphi| \mu_{1,\dots,2^{N-\ell}}, V_{1,\dots,N-\ell} )  &=&
\frac{i}{2} \sum_{i=1}^{2^{N-\ell}}  \mu^{(\ell)}_i 
\phi_{i}^{2}   \nonumber \\
&&\hspace*{-2.5cm}+ J^{(\ell)} \, W^{(N-\ell)}( \phi_{1,\dots,2^{N-\ell}}, V_{1,\dots,N-\ell} ) . \nonumber 
\end{eqnarray}
The renormalized parameters fulfill the recurrence
equations \cite{MonthusGarel}
\begin{align}
&\mu^{(\ell)}_i  = \frac{2 \mu^{(\ell-1)}_{2i-1} 
\mu^{(\ell-1)}_{2i} }{\mu^{(\ell-1)}_{2i-1} + \mu^{(\ell-1)}_{2i} }
+ 2 J^{(\ell-1)} \,, \label{eqmu} \\
&J^{(\ell)} = J 2^{-\ell(\alpha-1)} \,,
\label{eqJ}
\end{align}
where $i=1,\dots,2^{N-\ell}$ and $\ell = 1,\dots,N$. The initial
values $\{ \mu^{(0)}_i \}_{i=1,\dots,2^{N}}$ and $J^{(0)}$ are 
the parameters of the resolvent in the original model. Equations (\ref{eqmu}) 
and (\ref{eqJ}) hold for a single realization
of the random Hamiltonian $\mathcal{H}_{N}$ with a finite size $L=2^{N}$ and 
random potentials drawn from an arbitrary distribution $p(\varepsilon)$.

This procedure further provides a set of recursion
relations for the resolvent matrix elements. After performing $\ell=N$
changes of integration variables, we end up with a single site resolvent 
characterized by the renormalized parameter $\mu^{(N)}_{1}$. This is 
the initial condition for the iteration of the
resolvent recurrence equations
from $\ell =N$ to $\ell=1$, which finally 
restores $\{ G_{ij}^{(N)} \}$ in the original system.
For a numerical calculation of the average DOS and the IPR using
the diagonal elements $\{ G_{ii}^{(N)} \}$ obtained
from this procedure, we
refer the reader to Ref. [\onlinecite{Metz2013}]. 

Here we study the flow of the distribution $\mathcal{P}^{(\ell)}(\mu)$
of the random variables $\{ \mu^{(\ell)}_{i} \}$, obtained from the iteration of
Eq. (\ref{eqmu}).
Since  $\{ \mu^{(\ell)}_{i} \}$ are interpreted as renormalized random 
potentials, the distinction between localized and extended
states should be accompanied by a qualitative change
of the fixed-point distribution $\mathcal{P}^{(\infty)}(\mu)$
 in the limit $\eta \rightarrow 0$. Throughout the rest
of the paper we work directly
at $\eta=0$, such that $\{ \mu^{(\ell)}_{i} \}$ are real variables
and the statistical properties derived from Eq. (\ref{eqmu})
are valid for a finite system size $L$. In spite
of that, we will be interested
in the behaviour of $\mathcal{P}^{(\infty)}(\mu)$ when $L$ becomes
very large, which eventually leads to strong fluctuations
of $\{ \mu^{(\ell)}_{i} \}$ due to the presence of arbitrarily
small denominators in Eq. (\ref{eqmu}).
These unbounded fluctuations are suppressed by any
nonzero value of $\eta$, affecting the stability
of the different fixed-point distributions in a decisive way.
An analogous approach has been used in the  context of Levy 
random matrices, \cite{Bouchaud,Burda} where the resolvent matrix elements
are calculated directly at $\eta=0$.
The random potentials $\{  \varepsilon_i \}$
enter solely in the initial distribution $\mathcal{P}^{(0)}(\mu)$ and they constitute the unique 
source of randomness in the flow of $\mathcal{P}^{(\ell)}(\mu)$.
We expect that 
a fixed-point distribution $\mathcal{P}^{(\infty)}(\mu)$
is attained for finite values of $\ell$, provided $L$ is sufficiently
large.

The distribution $\mathcal{P}^{(\ell)}(\mu)$
can be computed analytically in two limiting 
situations.
In the pure model, 
where $p(\varepsilon) = \delta(\varepsilon)$, 
it is easy to show that 
\begin{equation}
\nonumber
\mathcal{P}^{(\ell)}(\mu) = \delta(\mu - E^{{\rm pure}}_{\ell} + E).
\end{equation}
By setting  $E=E^{{\rm pure}}_{\infty}$ and taking the limit
$\ell \rightarrow \infty$ we obtain the fixed-point distribution $\mathcal{P}^{(\infty)}_{p}(\mu) =\delta(\mu)$.

The second solvable case is represented
by a Cauchy distribution 
\begin{equation}
\nonumber
p(\varepsilon) = \frac{\gamma}{\pi(\gamma^2 + \varepsilon^2)}
\end{equation}
characterized by a scale parameter $\gamma > 0$ and a divergent variance.
In this case, one has \cite{MonthusGarel}
 \begin{equation}
 \mathcal{P}^{(\ell)}(\mu) = \frac{\gamma}{\pi  \left[ \gamma^2 + (\mu - E^{{\rm pure}}_{\ell} + E)^2 \right]}.
 \end{equation}
  Setting  once again $E=E^{{\rm pure}}_{\infty}$, in the
$\ell \rightarrow \infty$ limit we  obtain the fixed-point distribution
\begin{equation}
\mathcal{P}^{(\infty)}_{c}(\mu) = \frac{\gamma}{\pi(\gamma^2 + \mu^2)}.
\end{equation}

The convergence towards the stationary solution $\mathcal{P}^{(\infty)}_{p}(\mu)$ is naturally
interpreted as a signature of the extended phase, since the RRP 
do not fluctuate from site to site. Besides that, the band edge wavefunction
corresponding to $\mathcal{P}^{(\infty)}_{p}(\mu)$ uniformly spreads  throughout the whole system.
On the other hand, the strong
fluctuations of the RRP, due
to the 
Cauchy distribution $\mathcal{P}^{(\infty)}_{c}(\mu)$ and its
divergent variance, are
characteristic of the localized phase. 
We point out that spectral localization
has been proven in the whole range of parameters 
when $\{ \varepsilon_i \}_{i=1,\dots,L}$ are Cauchy distributed random variables. \cite{Molchanov1996}

The study of the pure model or of the initial Cauchy distribution
is less interesting, since  the extended or localized
fixed-points are stable in the whole parameter space, depending whether
we choose a distribution $p(\varepsilon)$ with
zero or infinite variance, respectively. The choice
of a distribution $p(\varepsilon)$ with a finite variance
will eventually lead to a competition for stability among 
$\mathcal{P}^{(\infty)}_{c}(\mu)$ and $\mathcal{P}^{(\infty)}_{p}(\mu)$.
For a distribution
$p(\varepsilon)$ with a finite variance, spectral localization has been proven
for $\alpha > 3/2$, \cite{Kr07} while numerical results for the
average IPR support the presence of extended
states in the same range of $\alpha$. \cite{Metz2013}

\section{Weak disorder expansion} \label{secperturb}

In order to perform an expansion of Eq. (\ref{eqmu}) in powers
of the disorder strength $W$, we rescale the random potentials
as $\varepsilon_i \rightarrow W \varepsilon_i$ and assume that
they are drawn from a distribution 
with $\langle \varepsilon_i \rangle_{\varepsilon} = 0$ and
$\langle \varepsilon_i \varepsilon_j \rangle_{\varepsilon} = \delta_{ij}$. 
We assume that $W/J \ll 1$ and, for the initial iteration steps, we expand Eq. (\ref{eqmu}) 
up to order $O(W^2)$, from which we derive the following expression for arbitrary $\ell$
\begin{eqnarray}
\mu_{i}^{(\ell)} &=& E_{\ell}^{{\rm pure}} - E + \frac{W}{2^{\ell}} \sum_{k=1}^{2^{\ell}} \varepsilon_{2^{\ell}i + 1 -k} 
\label{eqperturb}
\\
\nonumber 
&+& W^2 \sum_{p=1}^{\ell} \frac{1}{2^{p+\ell} \left(E - E^{{\rm pure}}_{p-1} \right)  } \sum_{r=1}^{2^{\ell-p}} 
 \left( \sum_{k=1}^{2^{p-1}}   \xi^{(\ell)}_{k,r,p}  \right)^2 ,
\end{eqnarray}
with  
\begin{equation}
\nonumber
\xi^{(\ell)}_{k,r,p} \equiv \varepsilon_{2^{\ell}i -(k-1) - (r-1)2^{p}} - \varepsilon_{2^{\ell}i -(k-1) - (r-1)2^{p}- 2^{p-1}} \, .
\end{equation}
From Eq. (\ref{eqperturb}) one can compute the average
\begin{eqnarray}
\langle \mu_{i}^{(\ell)} \rangle_{\varepsilon} &=& E_{\ell}^{{\rm pure}} - E + m_{\ell}(E) W^2 ,
\label{eqmean} \\
m_{\ell}(E) &=& \sum_{p=1}^{\ell} \frac{1}{2^{p} \left(E - E^{{\rm pure}}_{p-1} \right)  } ,
\end{eqnarray}
and the standard deviation
\begin{equation}
\Delta_{\ell} =   \sqrt{\langle (\mu_{i}^{(\ell)})^2 \rangle_{\varepsilon} -    \langle \mu_{i}^{(\ell)} \rangle_{\varepsilon}^2} = \frac{W}{2^{\ell/2}} \, ,
\label{eqvarperturb}
\end{equation}
in which we have retained terms up to $O(W^2)$. By calculating $\langle (\mu_{i}^{(\ell)})^3 \rangle_{\varepsilon}$
and $\langle (\mu_{i}^{(\ell)})^4 \rangle_{\varepsilon}$ one can check that  $\{ \mu^{(\ell)}_{i} \}$
are Gaussian distributed random variables,
independently of the details of  $p(\varepsilon)$.

The behavior of $m_{\infty}(E)$ determines whether the
perturbative expansion
is convergent or not. One immediately notes that $m_{\infty}(E)$ diverges whenever
we choose $E$ at one of the energies of the pure spectrum. 
This situation is trivial in the sense that the
eigenstates at $E = E_{p}^{{\rm pure}}$ ($p < \infty$) are localized for arbitrary
weak disorder.\cite{Metz2013}
The extended eigenstate which may remain stable for $W > 0$
is located at $E = E_{\infty}^{{\rm pure}}$.
In this case, the behaviour of $m_{\ell}(E_{\infty}^{{\rm pure}})$ for $\ell \rightarrow \infty$ depends 
on $\alpha$ according to
\begin{eqnarray}
\alpha &>& 2: \, m_{\ell}(E_{\infty}^{{\rm pure}}) \propto 2^{\ell(\alpha -2)} \xrightarrow{\ell \rightarrow\infty} \infty \nonumber \, , \\
\alpha &=& 2: \, m_{\ell}(E_{\infty}^{{\rm pure}}) \propto \ell \xrightarrow{\ell \rightarrow\infty} \infty \nonumber \, , \\
\alpha &<& 2: \, m_{\ell}(E_{\infty}^{{\rm pure}}) \xrightarrow{\ell \rightarrow\infty} \frac{1}{2 (E_{\infty}^{{\rm pure}} - E_{0}^{{\rm pure}}) (1 - 2^{\alpha -2} )  } \nonumber \,.
\end{eqnarray}
From the perturbation expansion it follows
that the delta peak fixed-point distribution becomes unstable
for arbitrary weak disorder as long as $\alpha \geq 2$.

Up to now we have been disregarding the conditions
of validity of the perturbative approach. Let us have a
closer look on this issue by making the following
change of variables 
\begin{equation}
\nu_{i}^{(\ell)} = \mu_{i}^{(\ell)} + E - E^{{\rm pure}}_{\ell}  \,, 
\end{equation}
which allows us to rewrite Eq. (\ref{eqmu}) as
follows
\begin{equation}
\nu_{i}^{(\ell)} = \frac{2 \nu_{2i-1}^{(\ell-1)} \nu_{2i}^{(\ell-1)} - (\nu_{2i-1}^{(\ell-1)} + \nu_{2i}^{(\ell-1)}) (E - E^{{\rm pure}}_{\ell-1})}
{\nu_{2i-1}^{(\ell-1)} +  \nu_{2i}^{(\ell-1)} - 2 (E - E^{{\rm pure}}_{\ell-1})} \,.
\label{eqflowA}
\end{equation}
From Eq. (\ref{eqflowA}) it is more straightforward to understand why 
perturbation might fail. 
The expansion of Eq. (\ref{eqflowA}) up to order $O(W^{2})$ 
is a good approximation throughout the whole renormalization flux provided that $|\nu_{2i-1}^{(\ell)} +  \nu_{2i}^{(\ell)}| \ll |2 (E - E^{{\rm pure}}_{\ell})|$.
If, on the contrary, $|\nu_{2i-1}^{(\ell)} +  \nu_{2i}^{(\ell)}| \approx |2 (E - E^{{\rm pure}}_{\ell})|$
for a certain $i$ and $\ell$, a small denominator arises in Eq. (\ref{eqflowA}), and 
the approximation given by Eq. (\ref{eqperturb}) breaks down 
for $\nu_{i}^{(\ell+1)}$. 
This resonance-like effect yields RRP 
with anomalous large magnitudes and we expect that the variance of
their distribution will exhibit an abrupt increase.

Although the failure of the perturbative results depends
crucially on the fluctuations of the RRP, we can estimate
the value of $\ell$ at which the perturbation breaks
down for
$E=E^{\rm pure}_{\infty}$. Let us assume that $W/J \ll 1$ and the
flow  
evolves according to perturbation in the
first iteration steps, since $E^{{\rm pure}}_{\infty} - E^{{\rm pure}}_{0} = O(1)$. As 
a consequence, keeping contributions up to order $O(W)$, $\{ \nu_{i}^{(\ell)} \}$ are Gaussian
distributed random variables with mean zero and standard
deviation $W/2^{\frac{\ell}{2}}$. The simplest approximation consists in treating all sites 
on the same footing by choosing $\nu_{i}^{(\ell)} = O(W/2^{\frac{\ell}{2}})\,\, \forall \, i $.
In this setting, perturbation fails for a value of
$l = l_{*}$ such that $W 2^{-\frac{\ell_{*}}{2}} = (E^{{\rm pure}}_{\infty} - E^{{\rm pure}}_{\ell_{*}})$, which
leads to
\begin{equation}
\frac{W}{J} = \frac{2^{-\ell_{*} (\alpha - \frac{3}{2}) + 1  }}{1 - 2^{1-\alpha} } \,.
\label{eqbreak}
\end{equation}
For $1 < \alpha < \frac{3}{2}$, there is no positive
value of $\ell_{*}$ which solves Eq. (\ref{eqbreak}), since the right
hand side diverges as a function of $\ell_{*}$ and $W/J \ll 1$ by construction.
For $\alpha >  \frac{3}{2}$, there is always a value of 
$\ell_{*}$ for which Eq. (\ref{eqbreak}) is fulfilled, since
the right hand side vanishes exponentially for increasing $\ell_{*}$. This value
is given by
\begin{equation}
\ell_{*} =\frac{\ln{ \Big[ \frac{2J}{W(1 - 2^{1-\alpha}) } \Big]}}{(\alpha-3/2) \ln 2}.
\label{eqbreak1}
\end{equation}

For fixed $W/J \ll 1$, $\ell_{*} \rightarrow \infty$ as $\alpha$ approaches $3/2$ from
above.
Equation (\ref{eqbreak1}) predicts that the 
perturbation expansion 
does not break down for $\alpha < 3/2$, such that 
the delta peak distribution is the only stationary
solution.

\section{Numerical results} \label{numsec}

In this section we discuss the numerical results
for the evolution of $\mathcal{P}^{(\ell)}(\mu)$ in connection
with the perturbative approach of the previous section.
The fact that Eq. (\ref{eqmu}) is defined for a finite system size $L$
represents a serious numerical restriction, since 
the total number of iteration steps
is limited by $L$. 
In order to overcome this issue,  a different route is followed in the numerical
calculation of $\mathcal{P}^{(\ell)}(\mu)$. The RRP $\mu^{(\ell)}_1,\dots,\mu^{(\ell)}_{2^{N-\ell}}$, {\it at
a given layer $\ell$}, are statistically independent random variables and their distribution  $\mathcal{P}^{(\ell)}(\mu)$
depends only upon  $\mathcal{P}^{(\ell-1)}(\mu)$. This allows us to implement
a population dynamics approach, which consists in parametrizing the distribution $\mathcal{P}^{(\ell)}(\mu)$ by a large
number $\mathcal{N}$ of stochastic variables representing instances of $\mu$.
To update $\mathcal{P}^{(\ell)}(\mu)$, we choose at random two
variables from the pool representing $\mathcal{P}^{(\ell-1)}(\mu)$, which are used to
update, according to
Eq. (\ref{eqmu}), a single variable extracted at random from the pool of layer $\ell$. This updating
rule is repeated 
until $\mathcal{P}^{(\ell)}(\mu)$ reaches a stationary form.
One expects that the statistical properties of the RRP converge
to a well-defined limit for large enough $\mathcal{N}$.
We remark that sample to sample fluctuations may arise
in the population dynamics algorithm due to finite values
of $\mathcal{N}$. In this sense, the population size $\mathcal{N}$ plays an
analogous role as $L$ in finite size calculations of Eq. (\ref{eqmu}).
For detailed
discussions of the population dynamics algorithm in the
context of spin-glasses and random matrices, we refer
the reader to Refs. [\onlinecite{ParisiMezard}] and [\onlinecite{Kuhn}], respectively.
\begin{figure}[t!]
\center
\scalebox{0.9}{\input{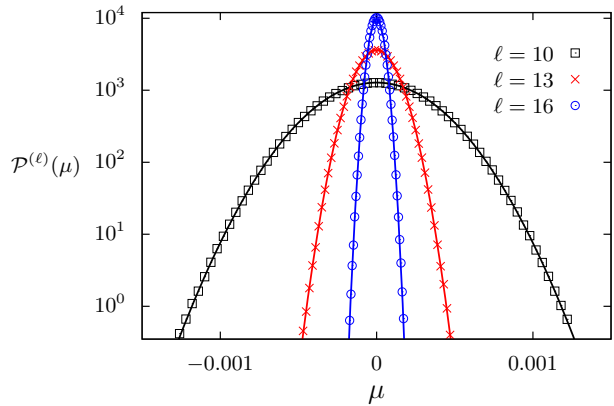}}
\caption{Numerical results  for the flow of the distribution
$\mathcal{P}^{(\ell)}(\mu)$ of the renormalized random 
potentials (taken away their mean value)  
 for $\alpha=1.25$ and  $E=E^{{\rm pure}}_{\infty}$. The
distribution $p(\varepsilon)$ has a Gaussian form, with mean zero and 
standard deviation $W=10^{-2}$. The solid lines show
Gaussian distributions with mean zero and standard deviations given 
by Eq. (\ref{eqvarperturb}).
The numerical data have been obtained through the population
dynamics algorithm with  $\mathcal{N}=10^{7}$ (see the main text).
}
\label{FlowDistr1}
\end{figure}

In all numerical results presented in this section, $p(\varepsilon)$ is a Gaussian distribution
with mean zero and standard deviation $W$. The initial values of $\mu^{(0)}_1,\dots,\mu^{(0)}_{\mathcal{N}}$
are generated according to $\mu^{(0)}_i = \varepsilon_i  + E_{0}^{\rm pure} - E$.
In addition, we 
set $J=1$ and restrict ourselves to the flow at the band edge of the
pure model, i.e., $E=E^{{\rm pure}}_{\infty}$. We are basically interested
in the behaviour of $\mathcal{P}^{(\infty)}(\mu)$ for different values of $\alpha$.

Figure \ref{FlowDistr1} shows the flow of 
$\mathcal{P}^{(\ell)}(\mu)$ for $W=10^{-2}$ and $\alpha = 1.25$.
The symbols are numerical results obtained from the
population dynamics method, while the solid lines are 
Gaussian distributions with mean zero and standard deviations 
for different values of $\ell$, given by Eq. (\ref{eqvarperturb}).
As can be seen, the agreement between the numerical and
the perturbation results is excellent for this value of $\alpha$, where
the delta peak is the only stable fixed-point distribution.
Figure \ref{FlowDistr1} illustrates the typical flow in the
extended phase: the initial Gaussian distribution $\mathcal{P}^{(0)}(\mu)$ shrinks
exponentially to a delta peak, characterizing the absence of 
fluctuations and the spatial homogeneity of the RRP.
\begin{figure}[t!]
\center
\scalebox{0.9}{\input{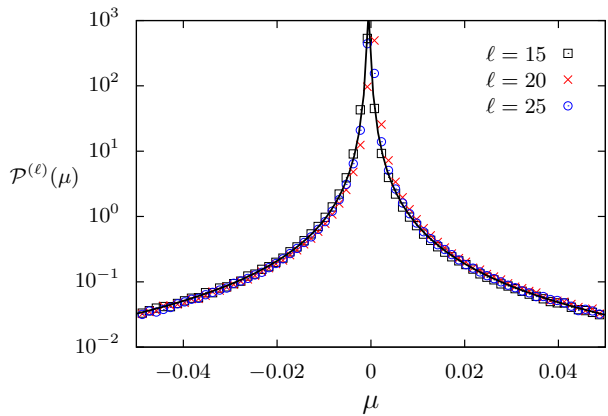}}
\caption{Numerical results for the flow of the distribution $P^{(\ell)}(\mu)$ of
the renormalized random potentials for $\alpha=2.25$ and  $E=E^{{\rm pure}}_{\infty}$. 
The distribution $p(\varepsilon)$ has a Gaussian form, with mean zero and 
standard deviation $W=10^{-2}$.
The solid line depicts
a Cauchy distribution with parameters taken from a fitting of the data
for $\ell=15$. The numerical data have been obtained through the population
dynamics algorithm with  $\mathcal{N}=10^{7}$ (see the main text).
}
\label{FlowDistr2}
\end{figure}

In figure \ref{FlowDistr2} we show the flow of
$\mathcal{P}^{(\ell)}(\mu)$ for $W=10^{-2}$, $\alpha = 2.25$ and
relatively large values of $\ell$. The perturbative
approach breaks down and
$\mathcal{P}^{(\ell)}(\mu)$
evolves to a Cauchy fixed-point distribution, as can be noticed
from the comparison between the population dynamics data (symbols) and
a Cauchy distribution obtained from a fitting of the
data for $\ell=15$ (solid line). This is the only stable fixed-point distribution
for this choice of $\alpha$, $E$ and $W$.
The presence of large, scale-free fluctuations
in the RRP typically yields localized eigenstates. 

\begin{figure}[t!]
\center
\scalebox{0.9}{\input{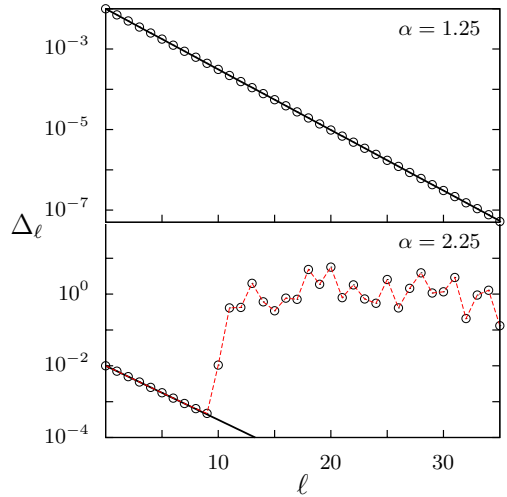}}
\caption{Numerical results for the flow of the standard deviation of $P^{(\ell)}(\mu)$
obtained from the population dynamics algorithm for
$\mathcal{N}=10^{7}$, $E=E^{{\rm pure}}_{\infty}$ and two values of $\alpha$. 
The distribution $p(\varepsilon)$ has a Gaussian form, with mean zero and 
standard deviation $W=10^{-2}$. The black solid line is the analytical
result of Eq. (\ref{eqvarperturb}), while the red dashed line is
just a guide.
}
\label{FlowStdmu}
\end{figure}

In order to clarify the breaking mechanism of the perturbative approach, 
figure \ref{FlowStdmu} exhibits the standard deviation $\Delta_{\ell}$
of $\mathcal{P}^{(\ell)}(\mu)$
for $\alpha=1.25$ and $\alpha=2.25$, corresponding to the data
in figures \ref{FlowDistr1} and \ref{FlowDistr2}, respectively. For $\alpha=1.25$, $\Delta_{\ell}$ 
vanishes exponentially as a function of
$\ell$ according to Eq. (\ref{eqvarperturb}). For $\alpha=2.25$, the flow of $\Delta_{\ell}$
is described by Eq. (\ref{eqvarperturb}) up to a certain $\ell$, at which
the presence of small denominators in Eq. (\ref{eqmu}) leads
to an abrupt increase of $\Delta_{\ell}$ by many orders of magnitude. This mechanism is responsible
for the emergence of strong fluctuations in the RRP, driving
the system to the Cauchy fixed-point distribution. In fact, the erratic behaviour of $\Delta_{\ell}$
for $\alpha=2.25$ and large values of $\ell$ is a signature that $\mathcal{P}^{(\ell)}(\mu)$ has evolved to
a Cauchy distribution.

\begin{figure}[t!]
\center
\scalebox{0.8}{\input{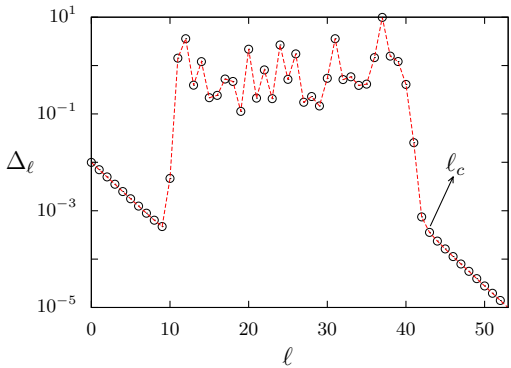}}
\caption{Numerical results for the flow of the standard deviation of $P^{(\ell)}(\mu)$
obtained from the population dynamics algorithm for
$\mathcal{N}=10^{7}$, $E=E^{{\rm pure}}_{\infty}$ and $\alpha=2.25$. 
The 
distribution $p(\varepsilon)$ has a Gaussian form, with mean zero and 
standard deviation $W=10^{-2}$. The value of $\ell$ where the Cauchy
fixed-point distribution becomes unstable is denoted by $\ell_c$.
The red dashed line is
just a guide.}
\label{fig1aspec}
\end{figure}

\begin{figure}[t!]
\scalebox{0.8}{\input{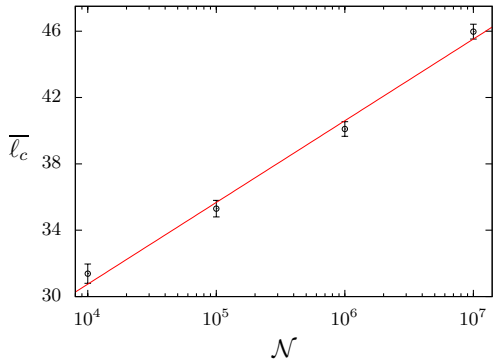}}
\caption{Average value of $\ell_c$ (see figure \ref{fig1aspec}) as a function of
the population size $\mathcal{N}$ for $\alpha=2.25$ and $E=E^{{\rm pure}}_{\infty}$.
The distribution $p(\varepsilon)$ has a Gaussian
form, with mean zero and standard deviation $W=10^{-2}$.
The average $\overline{\ell_c}$ is computed using $50$
independent runs of the population dynamics algorithm. The solid
line is the best fit  $\overline{\ell_c} = a + b \ln{\mathcal{N}}$ of the
data, with parameters $a = 11.1(1.3)$ and $b = 2.14(9)$.
}
\label{fig1app}
\end{figure}

For further larger values of $\ell$, we
eventually found that the 
Cauchy distribution usually becomes unstable and the parameters $\{ \mu^{(\ell)}_i \}$
flow back to a Gaussian distribution, until they finally reach
the delta peak distribution for $\ell \rightarrow \infty$. 
This effect is clearly illustrated in figure \ref{fig1aspec}, where
we present the standard deviation of $\mathcal{P}^{(\ell)}(\mu)$
up to $\ell =50$, for $\alpha=2.25$, $W=10^{-2}$ and $E=E^{{\rm pure}}_{\infty}$.
For intermediate values of $\ell$ the standard
deviation exhibits the erratic behaviour typical of the regime
where $\mathcal{P}^{(\ell)}(\mu)$ evolves to a Cauchy fixed-point distribution.
However, for $\ell \geq \ell_{c}$ the standard deviation presents once 
more the decay $\Delta_{\ell} \propto 2^{-\frac{\ell}{2}}$, reflecting the Gaussian
behaviour of $\mathcal{P}^{(\ell)}(\mu)$.
In order to probe the effect of the population size $\mathcal{N}$ on the 
stability of the Cauchy fixed-point, we have computed
the average of $\ell_{c}$
over a certain number of  independent runs of the population dynamics
algorithm. The outcome for $\alpha=2.25$, as a function 
of $\mathcal{N}$, is displayed in figure \ref{fig1app}.
The data
show that the mean value $\overline{\ell_{c}}$ diverges
as a logarithmic function of $\mathcal{N}$, strongly indicating that
the second Gaussian regime for larger $\ell$ is just an artifact
of the finite values of $\mathcal{N}$, and the
Cauchy distribution is the only stable solution for  
$\mathcal{N} \rightarrow \infty$ and large values of $\alpha$.

\begin{figure}[t!]
\center
\scalebox{0.8}{\input{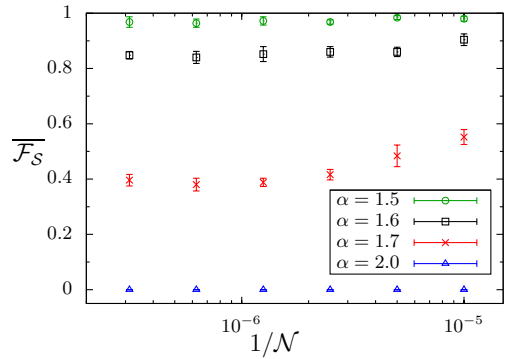}}
\caption{Average fraction of
runs of the population dynamics algorithm for which the standard deviation of
$\mathcal{P}^{(\ell)}(\mu)$ is given by Eq. (\ref{eqvarperturb}). 
The fraction $\mathcal{F}_{\mathcal{S}}$ is calculated
using $\mathcal{S} =50$ independent runs and $\overline{\mathcal{F}_{\mathcal{S}}}$
is computed by averaging $\mathcal{F}_{\mathcal{S}}$ over five independent data sets.
The initial configuration $\mu^{(0)}_{1},\dots,\mu^{(0)}_{\mathcal{N}}$ is
drawn from a Gaussian distribution with mean $E^{{\rm pure}}_{0} - E$ and
standard deviation $W=10^{-2}$. We have that $E=E^{{\rm pure}}_{\infty}$, and
the values of $\alpha$ are indicated on the figure.
}
\label{FracFixedPoint}
\end{figure}

In the population
dynamics method, the  sample to sample fluctuations
of the initial configuration $\mu^{(0)}_{1},\dots,\mu^{(0)}_{\mathcal{N}}$ may
have a significant impact on the stability of the stationary solutions.
The size of the basin of attraction of a given fixed-point distribution $\mathcal{P}^{(\infty)}(\mu)$
is proportional to the fraction of initial configurations that flow to $\mathcal{P}^{(\infty)}(\mu)$.
From a total of $\mathcal{S}$ independent
runs of the population dynamics algorithm, let us define $\mathcal{F}_{\mathcal{S}}$
as the fraction of runs in which the standard deviation of $\mathcal{P}^{(\ell)}(\mu)$
is given by Eq. (\ref{eqvarperturb}). 
We also define $\overline{\mathcal{F}_{\mathcal{S}}}$, i.e., the average value of
$\mathcal{F}_{\mathcal{S}}$ over different sets of samples of fixed
size $\mathcal{S}$. The quantity $\overline{\mathcal{F}_{\mathcal{S}}}$
provides a measure of the size of the basin of attraction of the
fixed-point delta peak distribution. 

We have computed the average fraction $\overline{\mathcal{F}_{\mathcal{S}}}$ over five
independent sets, each one containing $\mathcal{S}=50$  samples.
The behavior of $\overline{\mathcal{F}_{\mathcal{S}}}$ as a function of 
$\mathcal{N}$ is displayed 
in figure \ref{FracFixedPoint}, for $W=10^{-2}$, $E=E^{{\rm pure}}_{\infty}$
and different values of $\alpha$. 
As it can be
seen, in the regime of large $\mathcal{N}$ we have that
$\overline{\mathcal{F}_{\mathcal{S}}}=0$ 
for $\alpha = 2$, whereas $\overline{\mathcal{F}_{\mathcal{S}}} \simeq 1$
for  $\alpha = 1.5$. We have checked that $\overline{\mathcal{F}_{\mathcal{S}}} 
\rightarrow 1$ for fixed $\alpha = 1.5$ and decreasing $W$.
For $\alpha =1.6$ and $\alpha =1.7$, the fraction
$\overline{\mathcal{F}_{\mathcal{S}}}$ approaches a 
value $0 < \overline{\mathcal{F}_{\mathcal{S}}} < 1$ when $\mathcal{N} \gg 1$.
The numerical results on figure \ref{FracFixedPoint} strongly suggest that, for a certain interval
of values of $\alpha$, both the delta peak and the Cauchy distribution
are stable fixed-point distributions and the asymptotic behaviour
depends fundamentally on the initial configuration of the RRP.

\section{Final Remarks}
  
We have studied analytically and numerically the flow
of the distribution of the renormalized random potentials (RRP)
in the hierarchical Anderson model (HAM), characterized by 
a hopping energy
decaying as a power-law with exponent $\alpha$. More
specifically, we have focused
on the stability of the fixed-point distribution of the flow
at the upper spectral edge of the pure model, when a small
amount of on-site disorder is added to the system.
For large
values of $\alpha$ (short-range hopping), the RRP flow  to a Cauchy fixed-point
distribution, independently of their initial configuration.
This is consistent with the localization of all eigenstates
in low-dimensional tight-binding models with short-range hoppings. \cite{Gang4}
For small values of $\alpha$ (long-range hopping), the fluctuations of 
the RRP vanish exponentially and the flow converges to a
delta peak distribution. This is somehow consistent
with the  Wigner ensemble of random matrices, \cite{Wigner} where
the fully-connected infinite-range hoppings delocalize all eigenvectors.
In an intermediate range of $\alpha$ we have found that
the delta peak and the Cauchy distribution are both stable fixed-points, and
the asymptotic flow depends on the specific realization of 
the on-site disorder. 

Although
Eq. (\ref{eqbreak1}) implies that the perturbative approach
for the flow equations breaks 
down for $3/2 < \alpha < 2$, we have found numerically that
the RRP flow either to a delta peak 
or to a Cauchy distribution in this range of $\alpha$, depending on the 
initial configuration of the random parameters.
We point out that Eq. (\ref{eqbreak1}) has been
derived under a very crude assumption, namely that
{\it all} RRP are of the same order
of magnitude in the initial steps of the flow, which
amounts to neglect spatial fluctuations.
In spite of that, numerical and analytical
results seem to agree that for $\alpha < 3/2$ ($d_s > 4$)
the delta peak is the only fixed-point distribution, while
for $\alpha > 2$ ($d_s < 2$) the RRP always flow to the Cauchy
fixed-point distribution. 

Rigorous results have shown that for $\alpha > 3/2$ the HAM spectrum
contains solely a pure-point contribution, \cite{Kr08} whereas numerical
results for the inverse participation ratio support
the existence of extended eigenstates in the 
range $3/2 \leq \alpha \lesssim 2$, \cite{Metz2013} which coincides with 
the regime where the delta peak and the Cauchy distribution
coexist as stationary solutions of
the flow equations.
Overall, these results may indicate the presence of a mixed 
phase in the HAM, exhibiting features of localized and extended 
states. 
The study of the spatial decay of the 
wavefunctions and of the level-spacing distribution could
provide valuable information about the physical properties
in this intermediate regime of $\alpha$.
Analogous examples of mixed behaviour of localized
and extended features   
have been reported in the study of Levy random matrices  \cite{Bouchaud,Araujo}
and, more recently, in the Anderson model on the Bethe
lattice. \cite{Biroli2012} 
It would be interesting to investigate whether such 
unusual behaviour observed in the HAM is present close to the band edge
of high-dimensional tight-binding models with 
short-range hoppings.


\acknowledgments
We thank Vincent Sacksteder IV for interesting discussions
at an early stage of this work.
The research leading to these results has received funding from the European Research Council
(ERC) grant agreement No. 247328 (CriPheRaSy project),
from  the People Programme (Marie Curie Actions) of the European Union's Seventh Framework 
Programme FP7/2007-2013/ under REA grant agreement No. 290038  (NETADIS project) and 
from the Italian MIUR under the Basic
Research Investigation Fund FIRB2008 program, grant
No. RBFR08M3P4, and under the PRIN2010 program, grant code 2010HXAW77-008.

\bibliography{bibliography.bib}

\end{document}